\documentclass[1p,times,onecolumn]{elsarticle}

\usepackage{times,amsmath,epsfig}
\usepackage{subfigure}
\usepackage{cite}
\usepackage{comment}
\usepackage{amssymb}
\usepackage{graphicx}
\usepackage{gensymb}
\usepackage{listings}
\usepackage{epstopdf}
\usepackage{url}
\usepackage{balance}
\usepackage{harpoon}%
\usepackage{amsmath} 
\usepackage{color}
\usepackage{multirow}
\usepackage{afterpage}
\usepackage{algorithm}
\usepackage{verbatim}
\usepackage[noend]{algpseudocode}
\usepackage[table,xcdraw]{xcolor}
\usepackage{lipsum}
\usepackage{lineno}
\usepackage{soul}

\newfloat{program}{thp}{lop} 
\floatname{program}{Program}
\newcommand{\ie}{i.e.}

\journal{}

\begin{document}

\begin{frontmatter}

\title{SEGSys: A mapping system for segmentation analysis in energy}

\author[xiuli]{Xiufeng Liu\fnref{man}}
\address[man]{Department of Technology, Management and Economics, DTU}

\cortext[cor1]{Corresponding author}

\author[dtu]{Rongling Li\fnref{civil}}
\address[civil]{Department of Civil Engineering, DTU}

\author[yw]{Yi Wang\fnref{eth}}
\address[eth]{Department of Information Technology and Electrical Engineering, ETH}

\author[dtu]{Per Sieverts Nielsen\corref{cor1}\fnref{man}}
\ead{pernn@dtu.dk}

\begin{abstract}
Customer segmentation analysis can give valuable insights into the energy efficiency of residential buildings. This paper presents a mapping system, SEGSys that enables segmentation analysis at the individual and the neighborhood levels. SEGSys supports the online and offline classification of customers based on their daily consumption patterns and consumption intensity. It also supports the segmentation analysis according to the social characteristics of customers of individual households or neighborhoods, as well as spatial geometries. SEGSys uses a three-layer architecture to model the segmentation system, including the data layer, the service layer, and the presentation layer. The data layer models data into a star schema within a data warehouse, the service layer provides data service through a RESTful interface, and the presentation layer interacts with users through a visual map. This paper showcases the system on the segmentation analysis using an electricity consumption data set and validates the effectiveness of the system.
\end{abstract}


\end{frontmatter}
\date{}

\section{Introduction}
Smart meters have become widely used in recent years to collect fine-granular consumption data for electricity, heat, gas, and water. Moreover, smart metering of energy consumption has as part of the regulatory framework become mandatory in many countries, and in particularly applied in the residential sector. For example, European countries have made progress towards applying smart metering technologies for electricity use and heating use in the residential sector. As a result, a wealth of detailed energy consumption data is available, which provides a unique opportunity for energy operators to improve their operations and improve their services to their customers. The fine-grained energy consumption data can potentially be used to categorize load profiles and identify energy intense customers to tailor demand response and energy efficiency programs and improve energy savings. The data can be used to develop tools to identify customer groups related to load profiles and characteristics.

Modern energy systems require user-friendly tools to facilitate their own decision-making and ideally involve larger stakeholders in the decision making process. Regional energy operators often cannot effectively operate personalized services because they don't have the right tools. The tools need to be so sophisticated that the operator can tailor their services to specific customers. Although there are some tools available for energy efficiency decision-making, these tools, however, are usually closed and are offered as complete solutions without considering local specific needs. This is a problem because it doesn't provide sufficient emphasis on the social and economic aspects which are essential in user-centric energy services. Therefore, tailored tools are needed in order to improve energy services for both utilities and customers to take appropriate actions. This has been recognized as a growing need for user-friendly tools that can be used for non-professional users. These tools should integrate customers at the individual and local level (neighborhood) to improve flexibility, energy efficiency, and synergy effects. 

Data analytics is one of the tools which can be used in this decision making process. In particular customer segmentation analyses have become increasingly important in smart energy systems. It is a data analytical technique for demand-side management where customers are aggregated into multiple groups according to energy consumption characteristics and social characteristics of residents, including load patterns, load intensity, household data, and neighborhood data. The aim of this structured aggregation is to summarize a large number of customers into manageable subsets with similar characteristics \citep{motlagh}. Current literature, however, emphasizes much on the research of segmentation methods, such as self-organizing maps \citep{Hernandez}, k-means clustering \citep{cao} and its variants \citep{green,bahl}, hierarchy clustering methods \citep{merrick,nah} and so forth, while very few works emphasize visual analysis for customer segmentation. It is also important to note that geo-information system (GIS) data are becoming increasingly available. It is therefore possible to build a user-friendly GIS-based decision support tool for segmentation analyses, which is one of the most desirable features in energy system management and planning.

In this paper, we present a GIS-based system for customer segmentation based on load profiles, household characteristics and spatial information. The system is abbreviated as \emph{SEGSys} in the following. We introduce the BIRCH clustering algorithm \citep{birch} for customer segmentation analysis to distinguish heterogeneous daily energy consumption profiles because this algorithm allows  to detect anomalies related to irregular and scattered user behavior. The segmentation analysis consists of a descriptive modelling method based on hourly electricity consumption data, but the system is also applicable to other smart metering data such as heat, water, and gas. SEGSys comprises an online clustering module for the identification of typical energy load patterns and load intensity groups in real time; offline clustering based on physical geometries of neighborhoods; and offline clustering module for the classification according to household socioeconomic characteristics; and a front-end module for mapping and visualization. The proposed system can help utilities to make decisions according to different customer groups in an intuitive way. To the best of our knowledge, there are no previous studies that specifically segment customers by coupling different GIS data-sets and produce a physical segmentation on a map for demand-side energy consumption analysis.

More specifically, this paper makes the following contributions: 1) We implement a mapping system for energy segmentation analysis, and extensively explore the  segmentation analysis based on load patterns, consumption intensity, social characteristics, and neighborhood; 2) We present a mapping tool that allows users to interactively define the geometries at the individual and neighborhood levels of energy demand; 3) We present the in-database data mining and machine learning methods, as well as the mapping method on a visual map; 4) We showcase the segmentation of using an electricity consumption data set and validate its feasibility.

The remainder of the paper is constituted as follows: Section~\ref{sec:rev} reviews the related work. Section~\ref{sec:methods} presents the methods for segmentation analysis and the system implementation. Section~\ref{sec:casestudy} showcases the system in a real-world study. Section~\ref{sec:con} summarizes the paper and presents the future work.

\section{Literature review}
\label{sec:rev}
\subsection{Smart meter data analytic systems}

The increasing penetration of sensing and smart metering technology produce a large volume of energy consumption data. The digitalization of  smart energy systems provides opportunities for smart meter data analytics \citep{Gurung-2015,Zhou-2017}. Some systems or prototypes for smart meter data analytics have been found.  The dashboard, SmartD \citep{smartd}, and the system, SMAS \citep{smas},  were developed to analyze residential electricity consumption data, with the functionalities  including load profiling and pattern discovery. The latter also offers some advanced features such as customer segmentation and consumption disaggregation.  A smart meter data analytic pipeline was developed in \citep{liu2017}, which supports streamlining the entire data analytic process, including data acquisition, transformation, analysis and visualization. Ref \citep{liuy} developed a solution for big smart grid data management that uses Hive to process scalable consumption data in the cloud, but uses a relational database management system (RDBMS) to manage daily transaction data such as billing, user and organization information. These systems are orthogonal to this work, but SEGSys is dedicated to segmentation analysis, in combination with Geographic Information System (GIS) data, which aims at providing a  user-friendly interface for analysis and visualization.  Due to a variety of technologies available for smart meter data analysis, it is often difficult for users to select the right technologies for their needs. Ref \citep{wangy} gave a comprehensive review of smart meter data analytics with respect to applications, methodologies, and challenges. Ref \citep{liu2017} evaluated the technologies in the categories, including in-memory, in-database, parallelism in a distributed environment and parallelism through multi-threading. Based on the benchmarking results, PostgreSQL with MADlib \citep{madlib} is chosen to build the in-database analytic system SEGSys, due to its good performance and its simplicity of combining operational data and generating analytic views.

\subsection{Segmentation analysis}
The concept of customer segmentation was first developed in the 1950s by the marketing expert Wendell R. Smith, who classified customers by their value, demands, preference and other factors \citep{del}. In energy data management, segmentation analysis is an important technique for utilities to identify customer groups in order to provide targeting demand-respond programs and services. Much of customer segmentation research to date has been based on load profile studies. For example, refs \citep{McLoughlin-2015,Kwac-2014} extract features from load profiles of individual households and daily shapes, and then use the extracted features to segment customers through clustering. Customer groups are classified based on the most representative load profiles according to different day types (weekday, weekend and holiday) in \citep{Gianniou-2018,Carmo-2016} and different seasons (summer, winter, spring/autumn) in \citep{Fernandes-2017}. The variability of daily load profiles is used to select households with high potential for demand-response programs \citep{Gianniou-2018}.
Household characteristics are used as the features for clustering in \citep{Gianniou-2018,Carmo-2016,Beckel-2014,McLoughlin-2015}. Customers' insights, such as the groups classified according to their social characteristics or consumption behaviors, can help utilities make energy-saving recommendations and conduct effective energy campaigns \citep{Beckel-2014}. This is orthogonal to the work of segmenting customer groups based on sciodemographic factors in this paper. However, using  SEGSys, we segment customers according to not only  consumption load profiles, but also the social characteristic of households and the spatial characteristics of the neighborhood. SEGSys also has mapping function to visualize daily load patterns of dwellings from individual- to city-scale.

Clustering is the most used technique for segmentation analysis. It is an unsupervised machine learning method. It divides customers into several subgroups according to their similarity, which can be used to reveal the most typical load profiles \citep{Jain-1999}. The commonly used algorithms for clustering include the centroid-based methods such as k-means and k-medoids \citep{Jin-2017,Gianniou-2018,Fernandes-2017}, hierarchical clustering with agglomerative method \citep{Jin-2017,Fernandes-2017} and Self-Organizing Map (SOM) \citep{McLoughlin-2015,Beckel-2012}.  Ref \citep{Jin-2017} evaluated different clustering techniques using smart meter data and found that Centroid methods and hierarchical methods with Ward linkage perform better for segmentation. Ref. \citep{McLoughlin-2015} compared k-means, k-medoids and SOM methods based on a Davies-Bouldin (DB) validity index to identify the appropriate clustering methods and the corresponding number of clusters \citep{Davies-1979}. SOM shows consistently higher performance across a varying number of clusters. Regarding the sample size for clustering,  ref. \citep{Formann-1984} suggested that the minimal sample size should be no less than $2^k$ cases (k = the number of variables), preferably $5 \times {2^k}$. For small sample size data, hierarchical clustering algorithms are suitable \citep{Dolnicar-2002}. The clustering of the algorithms mentioned above is usually based on a large data set, and is performed offline because it is a time-consuming operation. For the online clustering for SEGSys, we choose the BIRCH algorithm \citep{birch} because it has a low memory footprint and a high efficiency. 

\subsection{Energy mapping}
The development of mapping tools for understanding where energy is consumed is important for energy systems. In \citep{triebe}, an energy flow map was developed to discover the improvement possibilities  in an energy flow. Ref \citep{fichera} evaluated energy demand of neighbourhoods and the most energy-intensive sectors of a city through mapping. Ref \citep{gango} assessed energy efficiency by mapping 129,635 existing buildings in Spain, and the results show that single-family houses consume on average more energy than individual dwellings. A model for evaluating energy efficiency of building blocks was also developed in \citep{dall}. The model uses data collected in energy audits of a random selection of buildings managed in a GIS platform. In \citep{asciono}, the energy building demand was assessed for both winter and summer, and mapped in urban areas to simplify the performance evaluation process. Ref \citep{gupta} presented a spatial approach to identify the neighborhoods for retrofits based on relative energy consumption and fuel poverty assessments. A GIS-based carbon mapping model is then used to shed energy use and saving potential at the household level. Therefore, energy mapping is an effective tool for problem identification and exploration of improvement potentials. In contrast to other works, this paper focuses on the customer segmentation mapping that can be used for providing personalized energy efficiency recommendations or services. In addition, both the online and offline segmentation are built into the system, and the online clustering algorithm supports anomaly detection in a timely manner.

\section{Methods}
\label{sec:methods}

\begin{figure*}[htp]
    \centering
    \includegraphics[width=0.9\textwidth]{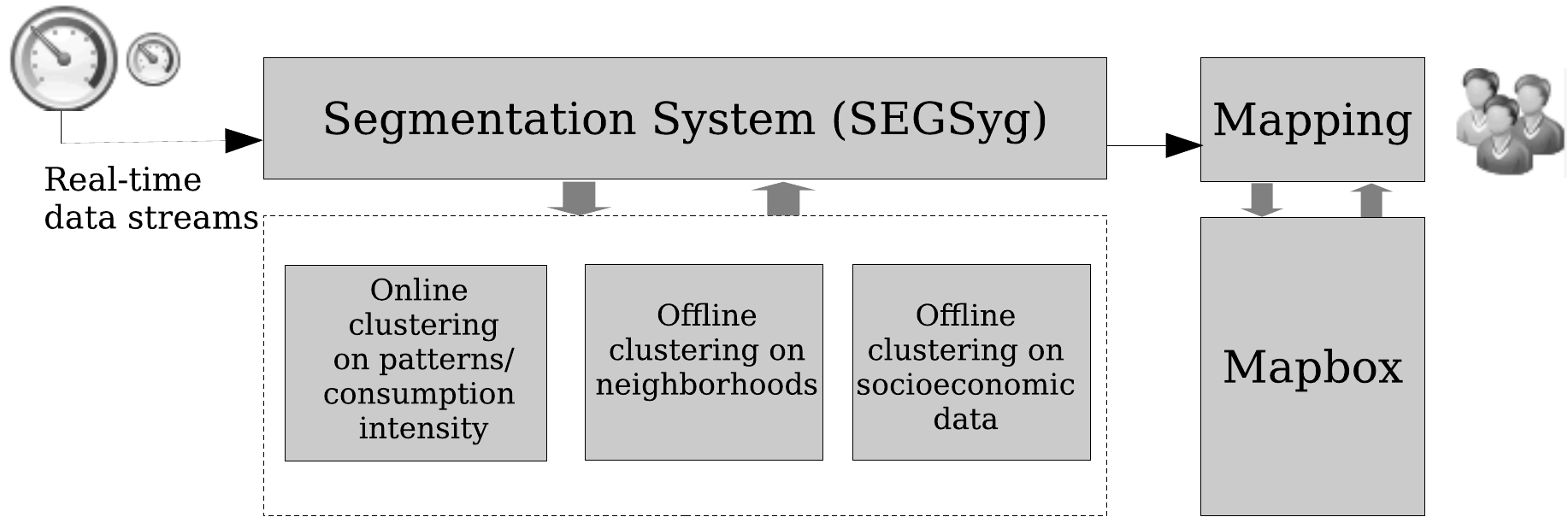}
    \caption{Overview of the segmentation analysis system for energy consumption}
    \label{fig:overview}
\end{figure*}

In this section, a complex customer segmentation system is developed to better investigate energy consumption and consumption behaviors for residential customers, as shown in Figure~\ref{fig:overview}. This system supports both online and offline customer segmentation analysis based on consumption intensity, patterns, geographic neighborhoods, and socio-economic data of customers, and the visualization of the analysis results using the method of mapping. More details about the segmentation and visualization method are provided in the following subsections.


\subsection{Segmentation analysis}
In this subsection, we present three segmentation methods for energy consumption analyses, including segmentation based on consumption patterns and consumption intensity using BIRCH clustering, segmentation based on neighborhood, and segmentation based on socio-demographic factors. Even though there are other segmentation methods for energy consumption such as methods based on based on lifestyles and market sectors \citep{marian2014}, we consider these three in this paper for the feasibility of our system implementation and the availability of data in general. 

\subsubsection{Segmentation based on consumption pattern and intensity}
Clustering is the core method used in this paper for the segmentation of energy consumption pattern and intensity.  In this paper, we employ the memory-based clustering algorithm, BIRCH \citep{birch}, to cluster data streams from smart meters. The reason that BIRCH is selected for the clustering in this paper are as follows: (1) BIRCH clustering scales well for big data, which has better performance than the other clustering algorithms including $k$-means and EM clustering. 
It is applicable to the applications that requires high performance or dealing with the large size of data, such as smart meter data, IoT data, and big image data. (2) It does not require the number of clusters as its inputs, which is different to $k$-means clustering algorithm. (3) BIRCH clustering algorithm can identify extreme values as anomalies.

BIRCH clustering algorithm uses a height-balanced tree, called {\em CF-tree}, as the data structure to store the condensed cluster features instead of all individual data points for clustering (see Figure~\ref{fig:cftree}). A node in the tree stores a cluster feature (CF) triple, $CF=(N, \overrightharp{LS}, \overrightharp{SS})$, where $N$ is the number of data points under this sub-tree, $\overrightharp{LS}$ is the linear sum of the $n$ data points, $\overrightharp{LS}=\sum_{i=1}^N \overrightharp{X}_i$; and $\overrightharp{SS}$ is the square sum of the $N$ data points, $\overrightharp{SS}=\sum_{i=1}^N (\overrightharp{X}_i)^2$. With the given $CF$, the clustering can proceed by calculating the measures without knowledge of the previous clustered data points. For example, the centroid $\overrightharp{C}$ and the  radius $R$ of the cluster can be calculated  by the following:
\begin{figure}[htp]
\vspace{-5pt}
\centering
\includegraphics[width=0.8\textwidth]{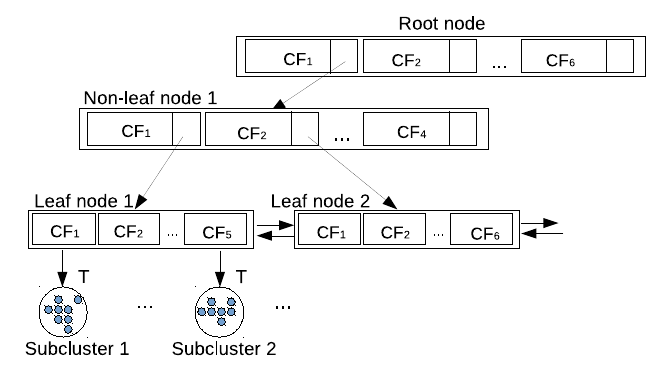}
\caption{The CF-tree, (L=3, B=6)}
\label{fig:cftree}
\end{figure}

\begin{equation}
	\overrightharp{C} = \frac{\sum_{i=1}^N \overrightharp{X}_i}{N} = \frac{\overrightharp{LS}}{N}  
\end{equation}
\begin{equation}
	R = \sqrt{\frac{\sum_{i=1}^N (\overrightharp{X}_i-\overrightharp{C})^2}{N}} = \sqrt{\frac{N\cdot\overrightharp{C}^2+\overrightharp{SS}-2\cdot\overrightharp{C}\cdot\overrightharp{LS}}{N}}
\end{equation}
For a data point $\overrightharp{X}_i$  and a cluster $CF_i=(N_i, \overrightharp{LS}_i, \overrightharp{SS}_i)$, the Euclidean distance of the point to the cluster centroid is denoted as  $D(\overrightharp{X}_i, CF_i)$. 
For two clusters, $CF_i=(N_i, \overrightharp{LS}_i, \overrightharp{SS}_i)$ and  $CF_j=(N_j, \overrightharp{LS}_j, \overrightharp{SS}_j)$, the Euclidean distance between the centroids is the distance of the two clusters, denoted as $D(CF_i, CF_j)$. If the two clusters are merged, the new centroid can be calculated as$(N_i+N_j, \overrightharp{LS}_i+\overrightharp{LS}_j,  \overrightharp{SS}_i+ \overrightharp{SS}_j)$.

The CF-tree has two control parameters, the branch factor $B$ and the threshold $T$. The number of entries of each node should not exceed $B$,  for example,  the entry number of each  node in Figure~\ref{fig:cftree} should be less than 6. When a new data point are added to the CF-tree, the data point starts from the root and recursively walks down the tree entering the nearest subcluster at the leaves.  When adding the new data point into the subcluster, the radius $R$ should not exceed the threshold value $T$, otherwise, a new cluster will be created. If the creation of a new cluster leads to more than $B$ child nodes of its parent, the parent will split, and the nodes above might also split recursively in order to maintain the tree balance. 

\begin{figure}[htp]
\vspace{-5pt}
\centering
\includegraphics[width=0.75\textwidth]{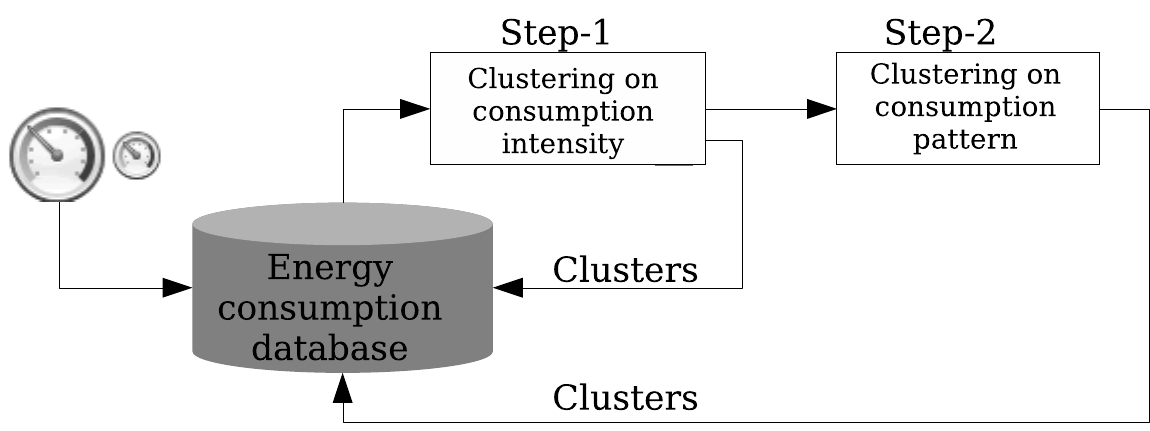}
\caption{Clustering on consumption intensity/patterns}
\label{fig:intensityclustering}
\end{figure}

According to the above discussion, CF-tree only keeps the aggregated data in the tree, and the size is much smaller than the original size of the data. The whole tree can be kept in memory for fast clustering. In this paper, we cluster daily load profile or pattern vectors,  $\{\overrightharp{X}_d|d=1, ..., N_d\}$, using BIRCH algorithm. The vector can be represented as  $\overrightharp{X}_d=<x_{0,d}, x_{1,d}, ..., x_{23,d}>$, where $x_{h, d}$ is the value of the hour $h$ for the day $d$. The clustering includes the following two steps (see Figure~\ref{fig:intensityclustering}). The first step is the segmentation based on load intensity, which is done for each household. The daily consumption can be classified into multiple categories according to the intensity. For the segmentation of consumption intensity, we also identify the extreme values by the BIRCH clustering, such as zero or extreme high values. The anomalies can be caused, for example, by meter defection, data transmission fault, theft or others. The detected anomalies are highlighted for user attentions when explore energy consumption history. 

The anomaly detection is based on the following. For a given threshold value $T$ in a clustering, when a new data point is added, the lower and upper limit of the distance of the data point to a cluster, $D$, can be computed using the parameters, $\alpha_1$ and $\alpha_2$, \ie, $T_L = \alpha_1 L$ and $T_U = \alpha_2 L$. The probability of a data point to be an anomaly can be calculated by

\begin{equation}
     prob = \begin{cases}
 0& \textbf{if } D\leq T_L \\ 
 1& \textbf{if } D\geq T_U \\ 
 \frac{D-T_L}{T_U-T_L}& \textbf{if } T_L < D < T_U
\end{cases}
\end{equation}
According to the probability, an anomaly can be decided, and users can set the parameters to define the sensitivity of the extreme value detection. To simplify the process, we set the same value for the two parameters in this paper such that a data point will be identified as an anomaly only when $D \geq T_U$.

The second step is the segmentation on the representative load patterns for all customers. This is done by clustering the normalized representative load profiles of all customers that are generated in the first step. The clustering is performed on the set of normalized representative load profiles,  $\{\overrightharp{X}^{'}_i |i = 1, ..., N_c \}$, where $\overrightharp{X}^{'}_i = <s_{0, i}, s_{1,i}, ..., s_{23,i}>$.  The normalization is formulated as follow:
\begin{equation}
s_{h, i} = \frac{x_{h,i}}{S_i}
\end{equation}
where $S_i$ is the sum of representative daily consumption profile of a customer $i$, \ie, $\sum_{h=0}^{23} x_{h,i}$.  Therefore, $\sum_{h=0}^{23} s_{h,i}=1$.

The following three metrics are used for the evaluation of BIRCH clustering performance, including entropy, the standard deviation of the cluster sizes, and estimated threshold, which is referred by ref  \citep{anthony2018}. 
\begin{equation}
E_k = - \frac{1}{N} \sum_{i=1}^k p(C_i)ln p(C_i)
\end{equation}

\begin{equation}
\delta_k = \sqrt{\frac{\sum_{i=1}^k(N(C_i) - \frac{1}{k}\sum_{j=1}^kN(C_j))}{n-1}}
\end{equation}
\begin{equation}
\theta_k = \frac{1}{N} \sum_{i=1}^k \frac{\sum_{X \in C_i}||X - C_i^0||}{||C_i^0||}
\end{equation}
where $N$ is the total number of vectors for clustering; $k$ is the number of resulting clusters; $p(\cdot)$ is the probability of a cluster; $C_i$ is a cluster of $i$;  $C_i^0$ is the centroid of a cluster of $i$;  $N(\cdot)$ is the cardinality of a cluster, and $||\cdot||$ is a 2-norm  distance.

\begin{table*}[t!]
{\small
\caption{The Postgis geometry functions used in SEGSys}
\label{tab:gisoperators}
\begin{tabular}{p{0.3cm}p{6.8cm}p{4.8cm}}
\hline
\textbf{No} &\textbf{Geometry function} & \textbf{Description} \\ \hline
1& {\em geometry} {\bf ST\_Intersection}({\em geometry A, geometry B})  & Returns that portion of geometry A and geometry B that is shared between the two geometries.  \\
2&{\em geometry} {\bf ST\_Union}({\em geometry[] g\_array})  & Returns the geometry that unions all the geometries in the array g\_array.   \\
3&{\em geometry} {\bf ST\_Difference}({\em geometry A, geometry B})  & Returns a geometry that represents that part of geometry A that does not intersect with geometry B.  \\ 
4&{\em boolean} {\bf ST\_Contains}({\em geometry A, geometry B})  & Returns TRUE if Geometry A contains Geometry B. It is used to find all the households within the neighborhood of A.  \\
5&{\em boolean} {\bf ST\_Equals}({\em geometry A, geometry B})  & Returns TRUE if the given geometry A equals the geometry B. It is used to retrieve an individual household. \\
\hline
\end{tabular}
}
\end{table*}

\subsubsection{Segmentation based on neighborhood}
Calculating energy consumption statistics based on neighborhood is useful for obtaining the value of each cell representing a specified neighborhood. For example, utilities can get an overview of the energy consumption for each neighborhood, and identify the neighborhoods with high and low energy demand so that they can improve energy management in the distribution. They can provide recommendations to houses with high consumption in the neighbourhood to improve energy efficiency. 

\begin{figure}[htp]
\centering
\includegraphics[width=0.55\textwidth]{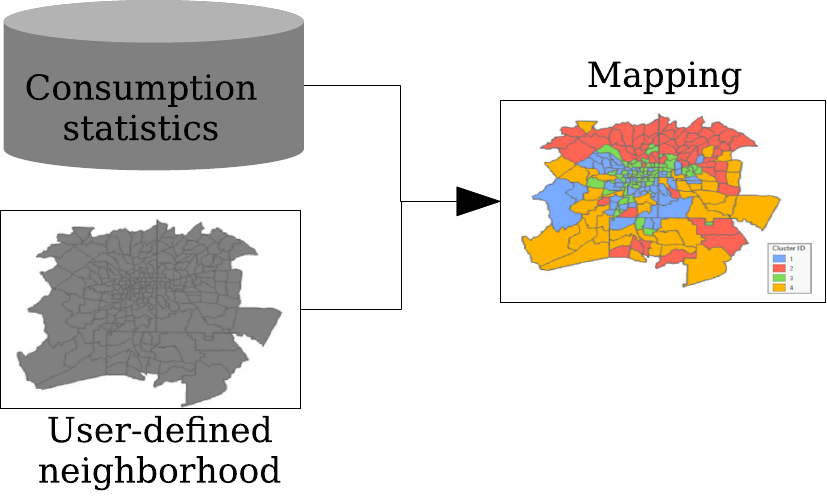}
\caption{GIS-neighborhood clustering}
\label{fig:gisclustering}
\end{figure}

In order to investigate energy demand at a district scale, energy consumption is segmented according to geographic neighborhoods or building blocks. With the geographic information of neighborhoods, the energy consumption statistics can be computed by an aggregation function, such as sum, mean  and percentile. The computed statistical data are visualized through the mapping on the geographic locations of neighborhoods  (see Figure~ \ref{fig:gisclustering}). In combination with the temporal characteristic of consumption data, it is possible to visualize these consumption statistic over time.  

\begin{figure}[htp]
\vspace{-5pt}
\includegraphics[width=0.45\textwidth]{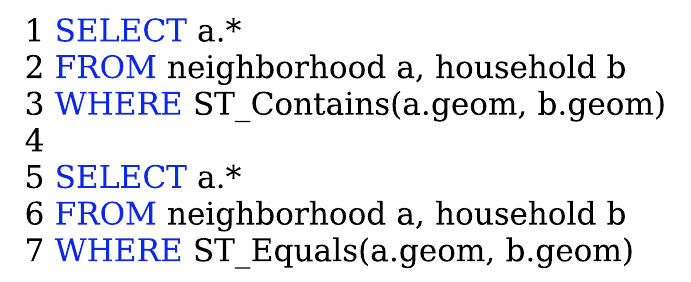}
\caption{SQL Statements using Postgis functions}
\label{fig:sqlstatement}
\end{figure}

The segmentation analysis for different neighborhoods was implemented using the GIS extension in PostgreSQL database, Postgis \citep{postgis}. Energy consumption statistics is computed by aggregating the values based on  neighborhood geometries. Postgis offers many operators and functions for geometries, including contain, equal, intersect, subtract, and others, which can be used to retrieve the households of interest based on the queries on geometries. Table~\ref{tab:gisoperators} lists five geometry functions used by SEGSys. The first three functions are used when the neighborhood geometries are created by the online mapping tool, which return the inter-, join- and subtracted geometry section, respectively. The latter two functions are used to query the households of interest, within a geometry boundary and a single geometry point, respectively. For example, the SQL statements in Figure~\ref{fig:sqlstatement} can return the households within a geometry area, and at a geometry point.

\subsubsection{Segmentation based on scio-demographic factors}
The socio-demographic factors related to residential energy consumption can be classified into two broad categories, household characteristics and dwelling characteristics. Household characteristics may include the size of family, age of the reference person, income, education, employment status, and so forth; while  dwelling characteristics may include dwelling type, building area, number of rooms, building age, etc. These factors may not be independent from each other. For example, building area is highly correlated to the size of family as a bigger family usually lives in a bigger apartment or house. A separate consideration of these correlated factors still makes sense as they might help to gain further insights into studying the impact on energy consumption \citep{marian2014}. There exist many studies on the effect of socio-demographic factors on energy consumption load profiles of households, which include the following. Ref \citep{kwac} measures the direct impact on household load profiles; Ref \citep{yohanis} investigates the impact of socio-demographic factors on maximum loads; and ref \citep{Schipper1989} studies the impact of socio-demographic factors on representative daily load profiles of heating consumption. The individual socio-demographic factors and their significance on energy consumption are summarized from previous work in Table~2. Therefore, it is interesting implement in  SEGSys which can explore the effect of individual factor or combined effect of multiple factors on energy consumption. 
\begin{figure}[t!]
\vspace{-5pt}
\centering
\includegraphics[width=0.95\textwidth]{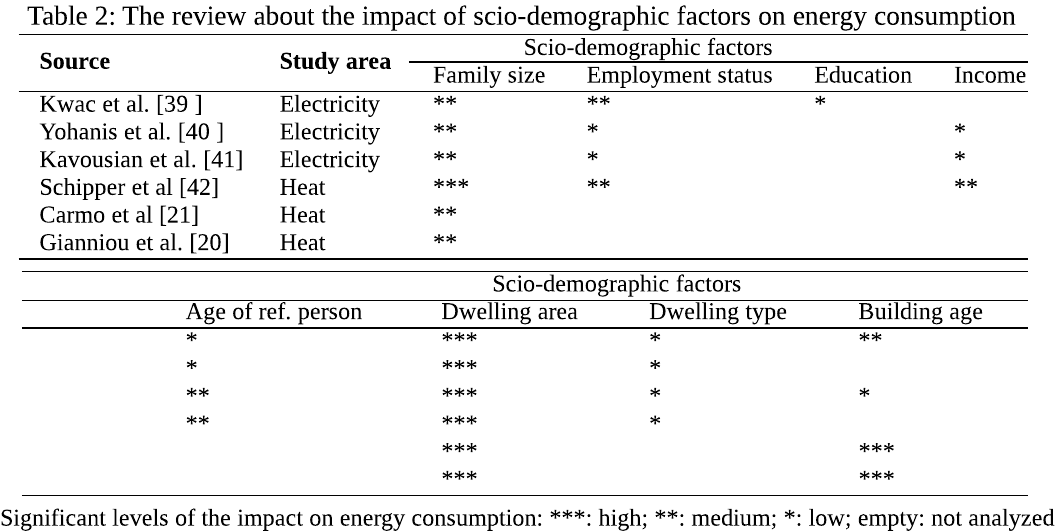}
\end{figure}
\begin{figure*}[t!]
    \centering
    \includegraphics[width=1\textwidth]{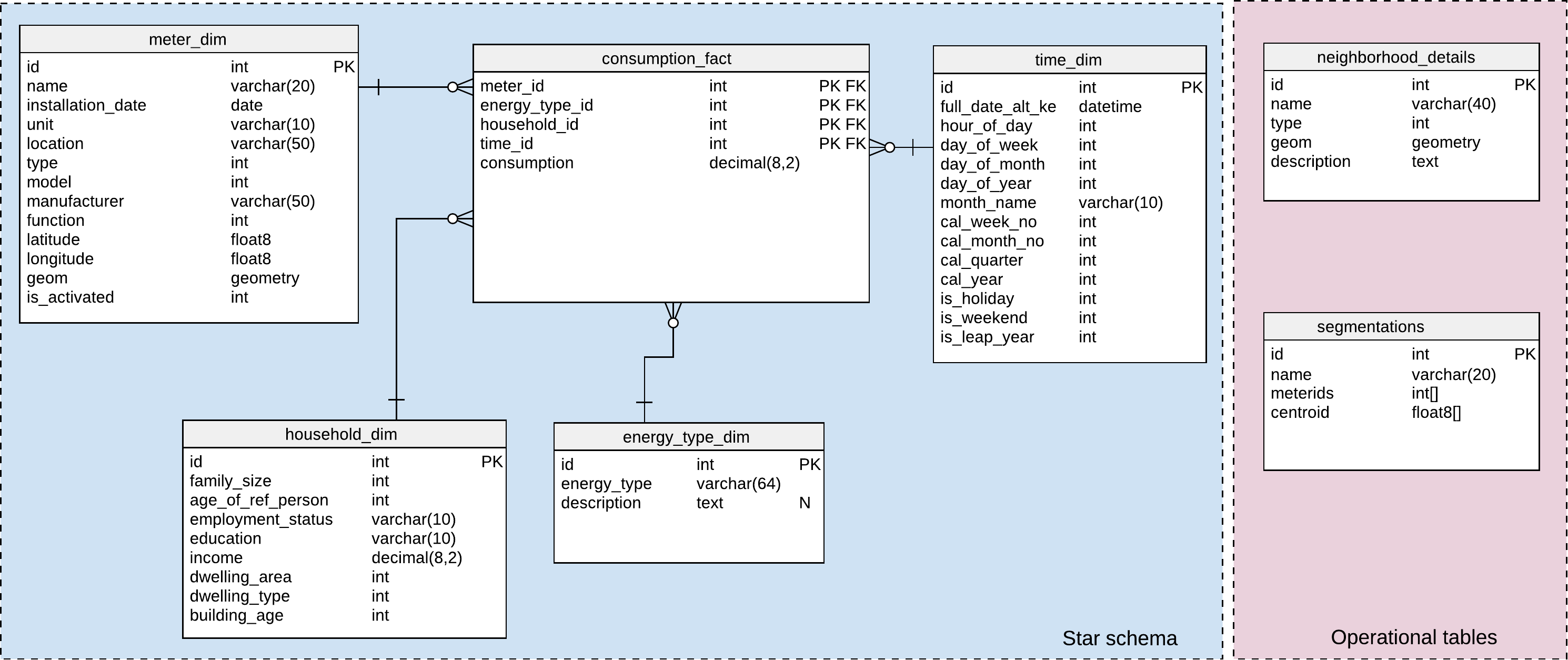}
    \caption{Overview of the segmentation analysis system for energy consumption}
    \label{fig:schema}
\end{figure*}

\subsection{System implementation}
In this section, we describe the implementation of SEGSys, including the design of data warehouse and the mapping system for visualization.

\subsubsection{Data warehouse}
We use the open source relational database management system, PostgreSQL, as the data warehouse system of SEGSys. In a real-world practice, there are two types of multidimensional models that can be used to structure data in a data warehouse, {\em star model}  and {\em snowflake model}. The star model has a demoralized schema, while the snowflake has a normalized schema. Therefore, the schema for a dimension in the snowflake model can be normalized into hierarchically structured tables. This can save more space, but  requires more table joins than in star model. In SEGSys, we choose the star model as the data warehouse schema designed to reduce possible table joins for better query performance. The logical model of the SEGSys data warehouse is shown in Figure~\ref{fig:schema}, which consists of a fact table (\texttt{consumption\_fact}) in the middle, and four dimension tables around the fact table. The fact table has a foreign-key reference on each dimension table. The fact table has the only metric, \texttt{consumption}, which stores the finest granular energy consumption values from data sources. This data warehouse model is designed for analytic queries for any  consumption type, such as electricity, water, and gas. The energy consumption type is modeled as a dimension table (\texttt{energy\_type\_dim}) and the corresponding meter as a separate dimension table (\texttt{meter\_dim}). The social characteristics of residents and the physical characteristics of a house or apartment are modeled as a household dimension table (\texttt{household\_dim}). We model different granular times into a flat time dimension table (\texttt{time\_dim}), instead of snowflake tables, to reduce table joins. In addition, the SEGSys data warehouse schema includes an operational table  (\texttt{neighborhood\_details}) for storing user-defined neighborhood geometries, and an operational table (\texttt{segmentations}) for storing segmenting customers. The operational tables support the operations including insert, update, and delete. The neighborhood geometries are generated by the online geometry generation tool in SEGSys. The segmentation information refers to the categorization of energy consumption intensity or patterns which can be generated by clustering or grouping the data. For example, the centroid of a cluster and its members are stored in an array-type column, allowing elements to be retrieved efficiently using a cluster identity as the primary key.

\subsubsection{Mapping and visualization}
SEGSys employs the open source mapping framework MapBox GL JS as its mapping system. Mapbox GL JS \citep{mapbox} can render vector based data as the map in the client side Web browser. A map can be made of multiple square tiles according to a tile coordinate system, and server in an image format at all zoom levels. The information to be displayed are organized into layers and features according to Mapbox Vector Tile specification, and each layer contains one or multiple features with geometrical information that describes the layer. There are three types of features for describing geometries, including Linestrings, Polygons and Points. SEGSys uses polygon and point geometries to describe energy consumption statistics and patterns at individual and neighborhood level, respectively. The geometries of individual households or neighborhood are defined by an online mapping tool in SEGSys, which represent the point or area of interest. User can pick the points or draw the polygons directly on a map. In addition, SEGSys supports importing shapefiles into the PostGIS database.

The visualization is to show the energy consumption statistics on a map using the MapBox. Energy consumption statistics or patterns were pre-computed in the data warehouse, and the output is retrieved from the PostGIS database by user queries. The data retrieved from PostGIS database can be expressed in different formats including XML, JSON, and GeoJSON. We choose GeoJSON as it is the mostly used format, which is well supported by Mapbox. The GeoJSON data are returend to the client, and rendered on a layer of the resulting map according to user-defined paint properties. A layer contains class-specific paint properties, which make it easy to change the map appearance. A layer can also set the optional min- and max-zoom properties that decide which zoom level the results should be rendered at.

\subsubsection{Web application}
The system of SEGSys is a web application implemented using the lightweight Python Web framework, Flask \citep{flask}. The system employs a service-oriented architecture. The server side is implemented as a RESTful web service provider that uses Flask framework to manage the routing  of incoming requests. The core of the services are built around objects, which  are stored in PostgreSQL tables in the data warehouse. Each row in a table represents an object for a target data point. The client  is implemented using the current web technologies including JavaScript, HTML5 and CSS. A request from the client is responded with JSON objects that contain the data queried from the data warehouse. A JSON object is human readable and parsed by the client-side JavaScript programs. The client uses the popular React JavaScript framework to coordinate user requests and interface updates in a responsive way on a single web page. The open source relational database system, PostgreSQL, is used as the underlying data warehouse, with the installation of the extension PostGIS for GIS data management. The in-database machine learning library, Apache MADlib \citep{madlib}, is used for data analysis within PostgreSQL. Our previous work  \citep{liu2017} has shown that the in-database analysis can achieve better performance as there is no network overhead for reading data out of database.
\begin{figure*}[t!]
\centering
\vspace{-5pt}
\includegraphics[width=0.95\textwidth]{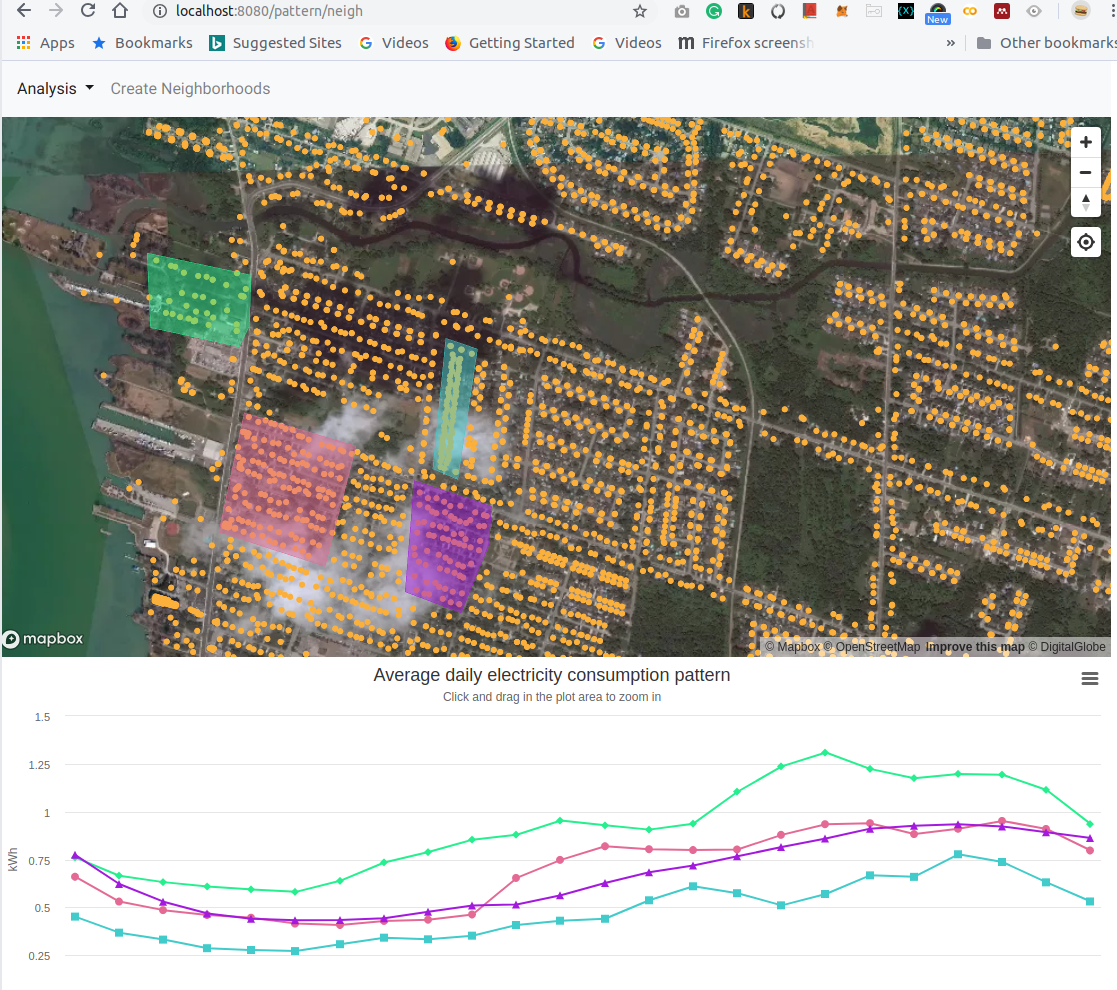}
\caption{User interface of SEGSys}
\label{fig:SEGSys}
\end{figure*}

\subsection{A case of study}
\label{sec:casestudy}

SEGSys is still under development. In particular, the data visualization is a complex task as we expect the customer segmentation information can be displayed on the map with given different conditions, for example, the selection of the social characterises of customers, the types of load patterns for clustering, the data range of temporal and the spatial dimensions. We now showcase our first version of SEGSys, which can segment customers based on the geometries of neighborhoods. Figure~\ref{fig:SEGSys} shows the screenshot of interactive interface of SEGSys. The top of the figure shows the geographical locations of each household (yellow dots). Users can select the neighborhoods to be compared with each other and then, for example, compare with the average daily consumption patterns of the selected neighborhoods. The bottom part of the figure shows the average daily consumption patterns corresponding to the selected neighborhoods in the above. The geometry of a neighborhood and the line of the load pattern are indicated in the same color.

For the segmentation by clustering, individual households will be shown by the dots in different colors. As the online clustering is supported, it is possible to display the transition of load pattern (if any) of a customer, for example, using a different color to dot a household over time. This will be an effective tool for utilities to learn about their customers, e.g., consumption behavior changes after some demand-respond programs. Further, the ``anomalies" of customers, e.g., the patterns or consumption intensity over a pre-set threshold value, can be highlighted on the map. These  will be our future work.



\section{Conclusions and future work}
\label{sec:con}
Segmentation analysis is an essential component of demand-side energy management. In this paper, we have implemented a segmentation analysis system, {\em SEGSys}, which  provides users with a decision support tool to intuitively monitor energy demand. We  presented the segmentation analysis with the approaches, including BIRCH clustering, customer-related spatial data analysis and social feature classification. We implemented the segmentation algorithms using in-database analysis technologies and designed a data warehouse for the segmentation system. We implemented the system as a web-based application with an integration of the mapping technology to enable a user-friendly visualization. We also provided a case study for the validation of the system.

There are many directions for future work. First, we will improve the user interface that allows users to select segmentation approaches and associated parameters. Second, we will adapt and test the design and implementation for the analysis of other energy sources such as heat, water and gas. Third, we will conduct a more comprehensive assessment or case studies to validate the system.

\section*{Acknowledgements}
This research was supported by the ClairCity project (http://www.claircity.eu) funded by the European Union's Horizon 2020 research and innovation programme (No.: 689289).

\bibliographystyle{elsarticle-num}

\flushleft

\end{document}